\documentclass[twocolumn,preprintnumbers,amsmath,amssymb,prb,aps,superscriptaddress]{revtex4}
\usepackage{txfonts}
\usepackage{pifont}
\usepackage{amssymb}
\usepackage{dcolumn}
\usepackage{amsmath}
\usepackage[dvips]{epsfig}

\makeatletter
\def\btt#1{\texttt{\@backslashchar#1}}%
\DeclareRobustCommand\bblash{\btt{\@backslashchar}}%
\makeatother

\begin{document}
\title{Spontaneous Fermi surface deformation in the three-band Hubbard model:
A variational Monte Carlo study}
\author{Xiao-Jun Zheng}
\affiliation{Key Laboratory of Materials Physics,
Institute of Solid State Physics, Chinese Academy of Sciences,
P. O. Box 1129, Hefei 230031, China}
\author{Zhong-Bing Huang}
\email{huangzb@hubu.edu.cn}
\affiliation{Faculty of Physics and Electronic Technology, Hubei University,
Wuhan 430062, China}
\affiliation{Beijing Computational Science Research Center,
Beijing 100084, China}
\author{Liang-Jian Zou}
\email{zou@theory.issp.ac.cn}
\affiliation{Key Laboratory of Materials Physics,
Institute of Solid State Physics, Chinese Academy of Sciences,
P. O. Box 1129, Hefei 230031, China}
\date{\today}

\begin{abstract}
We perform a variational Monte Carlo study on spontaneous $d$-wave form Fermi
surface deformation ($d$FSD) within the three-band Hubbard model. It is found that
the variational energy of a projected Fermi sea is lowered by introducing an
anisotropy between the hopping integrals along the x and y directions.
Our results show that the $d$FSD state has the strongest tendency at
half-filling in the absence of magnetism, and disappears as the hole concentration
increases to $n_h\approx 1.15$. This is qualitatively in agreement with the mean field
analysis and the exact diagonalization calculation for the one-band models, and provides
a qualitative explanation to the ``intra-unit-cell" electronic nematicity revealed by the
scanning tunneling microscopy. An analysis of the dependence of $d$FSD on
the parameters of the three-band model indicates that the copper on-site
Coulomb interaction, the nearest-neighbor copper-oxygen repulsion, and the charge-transfer
energy have a remarkable positive effect on $d$FSD.

PACS number(s):{74.20.Mn, 71.10.Li, 71.10.Fd}
\end{abstract}

\maketitle

\section{INTRODUCTION}
Hight temperature superconductivity is realized by doping charge carriers
into an antiferromagnetic insulator. In the underdoped region which
lies between hall-filling and optimal doping, competing orders
occur due to the interplay between the localized tendency arising
from the strong Coulomb repulsion and the delocalization due to
the mobilities of carriers. An interesting order tendency is the
nematic order,\cite{Kivelson1998} which breaks the four-fold
rotational symmetry of the underlying crystal but retains the
translational symmetry of the system. Recent neutron scattering
\cite{Hinkov2008} and Nernst effect\cite{Daou2010} measurements of
YBa$_2$Cu$_3$O$_y$ provided strong evidences for the existence of nematic
state in the underdoped cuprates. Moreover, a spectroscopic-imaging
scanning tunneling microscope (STM) investigation on Bi$_2$Sr$_2$CaCu$_2$O$_{8+\delta}$ \cite{Lawler2010}
suggested an intra-unit-cell (IUC) electronic nematic state which breaks
the ${90^ \circ }$ -rotational symmetry within every CuO$_2$ unit cell.

Theoretical studies on the two-dimensional $t$-$J$
\cite{Yamase2000,Miyanaga2006,Edegger2006} and Hubbard
\cite{Halboth2000,Valenzuela2001,Hankevych2002,Kampf2003,Neumayr2003}
models have reported a spontaneous $d$FSD instability in the
underdoped region: the Fermi surface (FS) expands along the
$k_{x}$ direction and shrinks along the $k_{y}$ direction or vice
versa. This order is considered to be generated by a forward
scattering of electrons close to the FS near $(0,\pi)$ and
$(\pi,0)$, which breaks the orientational symmetry but keeps the
translational symmetry of the system. From the symmetry point of
view such a state naturally leads to an electronic nematic order.
However, theoretical studies on the nematic order have so far mainly focused
on the one-band models or in the extreme limit of infinite
interactions~\cite{Kivelson2004}. A recent mean field analysis of a more
realistic model, i.e. the three-band Hubbard model~\cite{Emery1987}, showed that
the Coulomb interaction between next nearest-neighbor (NNN) oxygen atoms $V_{pp}$
plays a crucial role in the formation of the nematic order~\cite{Fischer2011},
which demonstrates that the interaction within the unit cell on the
CuO$_2$ plan is needed to be taken into account in order to show the delicate
mechanism and to depict a clear picture of the nematic order in the cuprates.

To gain a further insight into the nematic order in the cuprates,
we carry out a systematic variational Monte Carlo (VMC) study on
$d$FSD within the three-band Hubbard model.
Interestingly, we find that the electronic nematicity has already shown up
at $V_{pp}=0$, and becomes stronger with $V_{pp}$ increasing.
We also find that while both the copper on-site Coulomb interaction
$U_d$ and the charge-transfer energy ${\Delta_{ct}}$ have a remarkable
positive effect on $d$FSD, the NNN oxygen-oxygen hopping $t_{pp}$
has a negative effect. Our results are qualitatively in agreement with the
slave-boson mean field (SBMF) calculations~\cite{Yamase2000} over a wide
doping range, showing that $d$FSD has the strongest tendency around the
van Hove filling\cite{Halboth2000, Valenzuela2001, Neumayr2003, Yamase2005, Yamase2007}
(namely half-filing in our calculations) in the absence of magnetism.
Beyond that, we show that $d$FSD in the three-band model leads to an imbalance
in the hole densities of the neighboring oxygen sites, providing a consistent
theoretical explanation on the phenomena observed by STM in
Bi$_2$Sr$_2$CaCu$_2$O$_{8+\delta}$~\cite{Lawler2010}.

The paper is organized as follows: In Sec.II  we define the three-band Hubbard
model and outline the VMC scheme. Results from this numerical solution
and comparisons with previous studies are presented in Sec.III.
A conclusion is given in Sec.IV.

\section{Model Hamiltonian and Methods}
By considering the energetics and hybridizations for copper $3d_{x^2-y^2}$
orbital and oxygen $2p_x$ ($2p_y$) orbital, the kinetic part of the
three-band model reads,
\begin{eqnarray}
\label{eq:kinetic part}
 {H_0} &=& {\varepsilon _d}\sum\limits_{i,\sigma } {n_{i,\sigma }^d}
 + {\varepsilon _p}\sum\limits_{i,\sigma } {\sum\limits_\nu  {n_{i + \nu /2,\sigma }^p} }\nonumber\\
& &- {t_{pd}}\sum\limits_{i,\sigma } {\sum\limits_\nu  {\left( {d_{i,\sigma }^\dag {p_{i + \nu /2,\sigma }} + h.c.} \right)} }\nonumber\\
& &- {t_{pp}}\sum\limits_{i,\sigma } {\sum\limits_{\left\langle {\nu
,\nu '} \right\rangle } {\left( {p_{i + \nu /2,\sigma }^\dag {p_{i +
\nu'/2,\sigma }} + h.c.} \right)} },
\end{eqnarray}
where $d_{i,\sigma }^\dag$ and $p_{i+\nu /2,\sigma }^\dag$ create
a hole with spin $\sigma$ at the $i$th copper site and the
$i+\nu /2$th oxygen site, respectively, with $\nu=a_{x} (a_{y})$ being
the unit vectors along the x and y directions. $n_{i,\sigma }^d$ and
$n_{i + \nu /2,\sigma }^p$ are the corresponding number operators.
$t_{pd}$ and $t_{pp}$ denote the copper-oxygen and oxygen-oxygen hopping
integrals. $\left\langle {\nu ,\nu '} \right\rangle $ limits the sum over
NNN oxygen-oxygen lattice sites. We set ${\varepsilon _d}
\equiv 0$ on the copper site and introduce the charge-transfer energy
${\Delta _{ct}} ={\varepsilon _p} - {\varepsilon _d}$ to control the
relative Cu/O hole densities.

The interaction part of the model is given by,
\begin{eqnarray}
\label{eq:interaction part}
 H' &= &{U_d}\sum\limits_i {n_{i \uparrow }^dn_{i \downarrow }^d}
 + \frac{{{U_p}}}{2}\sum\limits_{i,\nu } {n_{i + \nu /2, \uparrow }^pn_{i + \nu /2, \downarrow }^p}\nonumber\\
 &&+ {V_{pd}}\sum\limits_{i,\nu } {\sum\limits_{\sigma ,\sigma '} {n_{i,\sigma }^dn_{i + \nu /2,\sigma '}^p} } \nonumber\\
 & & + {V_{pp}}\sum\limits_{i,\left\langle {\nu ,\nu '} \right\rangle } {\sum\limits_{\sigma ,\sigma '} {n_{i + \nu /2,\sigma }^pn_{i + \nu /2,\sigma '}^p} },
\end{eqnarray}
here $U_d$and $U_p$ are the on-site Coulomb interactions on the copper
and oxygen sites, respectively. $V_{pd}$ and $V_{pp}$ are the Coulomb
repulsions between nearest-neighbor (NN) copper-oxygen sites and
NNN oxygen-oxygen sites, respectively.
In the following, the parameters within CuO$_2$ plane are
measured in the unit of $t_{pd}$.

We use the following variational wave function:
\begin{eqnarray}
\label{eq:wave function} \left| \psi  \right\rangle  =
{P_G}\prod\limits_{\left| k \right| \le {k_F},\sigma } {\alpha
_{k\sigma }^\dag \left| 0 \right\rangle },
\end{eqnarray}
where ${P_G} = {P_d}{P_p}{P_{pd}}{P_{pp}}$ is the projection operator,
which contains four  parts: 1. ${P_d} = \prod\limits_i {\left( {1 - \left( {1 -
{g_d}} \right)n_{i \uparrow }^dn_{i \downarrow }^d} \right)} $
modifies the double occupancy on the Cu site; 2. ${P_p} = \prod\limits_{i,v}
{\left( {1 - \left( {1 - {g_p}} \right)n_{i + \nu /2, \uparrow }^pn_{i + \nu /2,
\downarrow }^p} \right)} $ modifies the double occupancy on the O site;
3. ${P_{pd}}= \prod\limits_{i,v} {{g_{pd}}^{n_i^dn_{i + \nu /2}^p}} $
adjusts the charge occupancies on the NN copper-oxygen sites; 4. ${P_{pp}} =
\prod\limits_{i,\left\langle {\nu ,\nu '} \right\rangle }
{{g_{pp}}^{n_{i + \nu /2}^pn_{i + \nu '/2}^p}} $ adjusts the charge
occupancies on the NNN oxygen-oxygen sites. Here $g_{d}$, $g_{p}$, $g_{pd}$, and $g_{pp}$
stand for four variational  parameters in the range from 0 to 1, which will
be optimized during the Monte Carlo calculations.

In Eq.~(\ref{eq:wave function}), the single-particle operator $\alpha _{k\sigma }^\dag$
is given by a linear combination of $d_{k\sigma}^\dag $, $p_{xk\sigma }^\dag $,
and $p_{yk\sigma }^\dag $, which are the Fourier transformations of $d-$ and
$p-$ hole operators:
\begin{eqnarray}
\label{eq:Fourier}
 d_{k,\sigma }^\dag &=& \frac{1}{{\sqrt N }}\sum\limits_i {d_{i,\sigma }^\dag {e^{ - ik{r_i}}}},  \\
 p_{vk,\sigma }^\dag  &=& \frac{1}{{\sqrt N }}\sum\limits_i {p_{i + \nu /2,\sigma }^\dag {e^{ - ik{r_{i + \nu /2}}}}},
 \end{eqnarray}
where $N$ is the total number of unit cells.
The coefficients in $\alpha _{k\sigma }^\dag$ are obtained through
diagonalizing the following matrix,
\begin{eqnarray}
\label{eq:Heff} &&{H_{eff}} =\nonumber\\
&& \left( {\begin{array}{*{10}{c}}
   {{\Delta _{eff}}} & { - 4{t_{pp}}\cos \frac{{{k_x}}}{2}\cos \frac{{{k_y}}}{2}} & { - 2{t_{pd,x}}\cos \frac{{{k_x}}}{2}}  \\
   { - 4{t_{pp}}\cos \frac{{{k_x}}}{2}\cos \frac{{{k_y}}}{2}} & {{\Delta _{eff}}} & { - 2{t_{pd,y}}\cos \frac{{{k_y}}}{2}}  \\
   { - 2{t_{pd,x}}\cos \frac{{{k_x}}}{2}} & { - 2{t_{pd,y}}\cos \frac{{{k_y}}}{2}} & 0  \nonumber\\
\end{array}} \right),\\
 \end{eqnarray}
here ${\Delta _{eff}}$ is a variational parameter acting as an
effective charge-transfer energy. Introducing an asymmetry parameter
${\delta _{{\mathop{\rm var}} }}$ between the hopping integrals
along $x$ and $y$ direction, we have
\begin{eqnarray}
\label{eq:Hopping}
{t_{pd,x}} = {t_{pd}} - {\delta _{{\mathop{\rm var}} }},\\
{t_{pd,y}} = {t_{pd}} + {\delta _{{\mathop{\rm var}} }}.
\end{eqnarray}
When $\delta _{{\mathop{\rm var}} }$ is finite, the shape of FS
is deformed. We evaluated the expectation value of total energy
for the variational wave-function by employing Monte Carlo method
\cite{Gros1988,Yokoyama1988,Yokoyama1996,Yamaji1998,Becca2000,Paramekanti2004}.
The four projection parameters $g_{d}$, $g_{p}$, $g_{pd}$, and $g_{pp}$,
as well as the effective charge-transfer energy $\Delta _{eff}$ and
the asymmetry parameter $\delta _{{\mathop{\rm var}} }$, are optimized
to obtain the minimal energy. During the optimization, a quasi-Newton method
combined with the fixed sampling method\cite{Ceperley1977,Umrigar1988} is used. The calculations
have been done for square lattices with periodic and antiperiodic
boundary conditions along the $x$ and $y$ direction, respectively.
These boundary conditions are used to avoid the degenerate states at
the Fermi surface. Typically, $10^7$ Monte Carlo steps are performed
for each set of variational parameters. The resulting statistical
errors are given in the relevant figures by error bars.

\section{NUMERICAL RESULTS AND DISCUSSIONS}
In the present paper, we take the typical values of the copper oxide
parameters. Setting the copper-oxygen hopping $t_{pd}\equiv 1$, we have
$U_{d} = 8.0$, $U_{p} = 3.0$, $V_{pd} = 1.0$, and $\Delta_{ct} = 3.0$,
according to the constrained density-functional
calculations~\cite{Hybertsen1989}. $t_{pp}$ is expected to have a
negative effect on $d$FSD, so we will firstly focus on the case of
$t_{pp}$=0.0 and the order dependence on $t_{pp}$ will be examined
alone later on. The results presented below are obtained for these
typical parameters, except where explicitly noted otherwise.

\begin{figure}[htbp]
\centering \setlength{\abovecaptionskip}{2pt}
\setlength{\belowcaptionskip}{4pt}
\includegraphics[angle=0, width=0.98 \columnwidth]{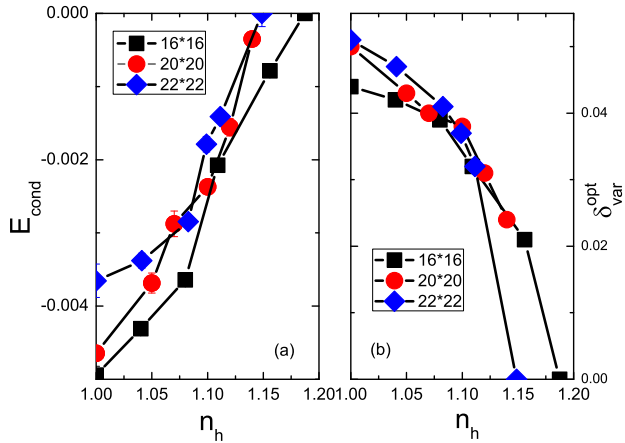}
\caption{(Color online) Condensation energy $E_{cond}$ (a) and optimized value of $\delta
_{{\mathop{\rm var}} }$ (b) as a function of hole density $n_h$ on
the 16$\times$16, 20$\times$20, and 22$\times$22 lattices.}
\label{fig:E_tvar_dp}
\end{figure}

Firstly, in Fig.\ref{fig:E_tvar_dp} (a), we present the doping dependence of the
condensation energy per unit cell ($E_{cond} =
\left[ {E\left({{\delta^{opt}_{{\mathop{\rm var}} }}} \right) - E(0)} \right]/N$, with
$\delta^{opt}_{\mathop{\rm var}}$ being the optimized asymmetry parameter)
on the square lattices of the sizes of 16$\times$16,
20$\times$20, and 22$\times$22. One can observe that for all the three
lattices, the absolute value of $E_{cond}$ decreases continuously from half-filling and
vanishes at $n_{h}\sim 1.15$. This behavior is consistent with the slave-boson
mean-field (SBMF) analysis for the $t-J$ model, indicating that the growing forward
scattering with decreasing hole density has a positive contribution to
the formation of the $d$FSD state. Fig.\ref{fig:E_tvar_dp} (b) shows the
dependence of $\delta^{opt}_{{\mathop{\rm var}} }$ on the hole density.
There is an obvious tendency that $\delta^{opt}_{{\mathop{\rm var}} }$ decreases
to zero at $n_h\sim 1.15$ in the same way as the condensation energy,
suggesting that $d$FSD only occurs in the underdoped region.

\begin{figure}[htbp]
\centering \setlength{\abovecaptionskip}{2pt}
\setlength{\belowcaptionskip}{4pt}
\includegraphics[angle=0, width=1.05 \columnwidth]{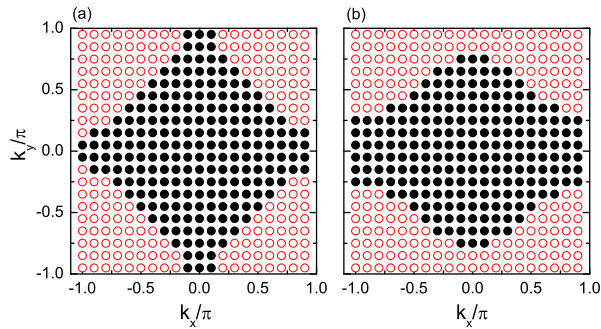}
\caption{(Color online) Fermi surfaces in the normal state ($\delta _{{\mathop{\rm
var}}}$=0.0) (a) and the $d$FSD state ($\delta _{{\mathop{\rm
var}}}$=0.062) (b) at the hole density $n_h$=1.05. Filled (empty)
sites indicate the occupied (unoccupied)
$k$ points.} \label{fig:2020dp105FS}
\end{figure}

We present the isotropic ($\delta _{{\mathop{\rm
var}}}$=0.0) and distorted ($\delta _{{\mathop{\rm
var}}}$=0.062) FS's at $n_h$=1.05 for
a 20$\times$20 square lattice in Fig.\ref{fig:2020dp105FS}. It is shown
that the four fold rotational symmetry of FS is reduced to two fold
in the distorted ground state. The FS expands along the $k_{x}$ direction
and shrinks along the $k_{y}$ direction as the optimized $\delta _{{\mathop{\rm var}} }$
takes a positive value. We learn from Fig.\ref{fig:2020dp105FS} that the finite size
effect brings some difficulties in determining the optimal value of $\delta
_{{\mathop{\rm var}} }$ in our calculations: when the FS is distorted by a
nonzero $\delta _{{\mathop{\rm var}}}$, the states near (0,$\pi$) or
($\pi$,0) undergo a discontinuous change from unoccupied to occupied or
vice versa. This kind of discontinuity introduces an uncertainty of the
optimal value of $\delta_{{\mathop{\rm var}} }$, which is estimated about
0.01$\sim$0.02 for our studied systems.

From Fig.\ref{fig:E_tvar_dp}, it can be seen that the $d$FSD state is stablest at
half-filling. However, strong $(\pi,~\pi)$ scattering around half-filling may favor
an antiferromagnetic (AFM) state, instead of the $d$FSD state.
To clarify this issue, we carried out VMC calculations for the projected
AFM state at half-filling, and the corresponding condensation energy
per unit cell is $-0.2185$ and $-0.2266$ on the 16$\times$16 and 20$\times$20 lattices,
respectively. Two order lower of the condensation energy in the AFM state
compared to the $d$FSD state demonstrates that magnetic ordering is
actually the strongest instability close to half-filling.
With the increase of doping concentration, the AFM order quickly disappears in
doped cuprates, and the ground state could be well described by a wave function
without long-range magnetic ordering, such as the one in Eq.~(\ref{eq:wave function}).

Another long-range order that may compete with $d$FSD is the $d$-wave superconductivity,
which is induced by AFM spin fluctuations. According to the VMC study of the one-band
$t$-$J$ model~\cite{Edegger2006}, the $d$FSD instability is overwhelmed by the $d$-wave
superconductivity. However, in the case of the three-band Hubbard model,
a VMC study~\cite{Yanagisawa2001} showed that the superconducting condensation energy
is about 10$^{-3}$, which is comparable to the energy gain in the $d$FSD state by our
calculations. This indicates that the $d$FSD instability is as strong as the
superconducting instability in the three-band Hubbard model. A further study is
needed to clarify one remaining question, whether both $d$FSD and $d$-wave
superconductivity can coexist in the multi-band model.

\begin{figure}[htbp]
\centering \setlength{\abovecaptionskip}{2pt}
\setlength{\belowcaptionskip}{4pt}
\includegraphics[angle=0, width=1.05 \columnwidth]{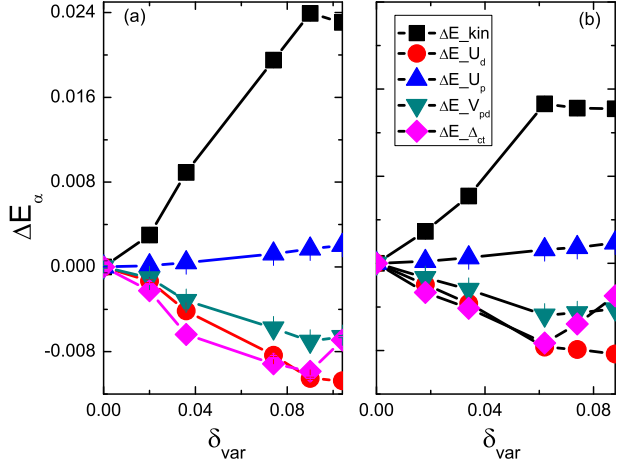}
\caption{(Color online) Different energy changes as a function of $\delta _{{\mathop{\rm
var}}}$ on the 20$\times$20 lattice. (a) $n_h$=1, (b) $n_h$=1.05. }
\label{fig:Ud=8detail}
\end{figure}

To understand the physical origin for the formation of $d$FSD,
we present different energy changes $\triangle
E_{\alpha} =(E_{\alpha}(\delta _{{\mathop{\rm var}}})-E_{\alpha}(0))/N$
as a function of $\delta _{{\mathop{\rm var}}}$, with $\alpha$ representing
different components of the Hamiltonian in
Fig.~\ref{fig:Ud=8detail}. We observe that while
$\triangle E\_kin$ and $\triangle E\_U_p$ take positive values as
the system changes into an anisotropic state, the other three components
$\triangle E\_U_d$, $\triangle E\_V_{pd}$, and $\triangle E\_\Delta _{ct}$
take negative values in the studied parameter region, which lower the total
energy of the system, and lead to a minimum at a certain value of $\delta
_{{\mathop{\rm var}}}$. This reveals that the $d$FSD state is the result
of delicate competition between the localized tendency introduced by $U_d$,
$V_{pd}$, $\Delta _{ct}$ and the delocalization due to the kinetic term of
carriers. One can notice that the dependence of different energy changes
on $\delta _{{\mathop{\rm var}}}$ shows similar behavior both at half-filling
($n_{h}=1$) and at a finite hole doping density ($n_{h}=1.05$). This is
rather different from the VMC study of the $t$-$J$ model at
half-filling~\cite{Miyanaga2006}, where the kinetic energy is exact zero and
the $J$ induced superexchange energy is found to increase with $\delta _{{\mathop{\rm var}}}$ 
increasing, resulting in the vanishing of condensation
energy at half-filling.

\begin{figure}[htbp]
\centering \setlength{\abovecaptionskip}{2pt}
\setlength{\belowcaptionskip}{4pt}
\includegraphics[angle=0, width=0.98 \columnwidth]{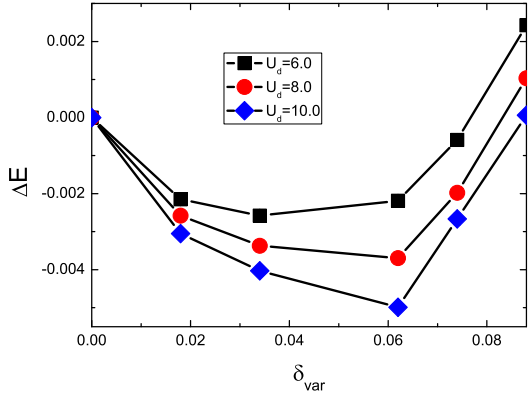}
\caption{(Color online) Total energy change as a function of $\delta _{{\mathop{\rm
var}}}$ for different values of $U_d$. Lattice = 20$\times$20 and
$n_h$=1.05.} \label{fig:2020dp105Ud}
\end{figure}

\begin{figure}[htbp]
\centering \setlength{\abovecaptionskip}{2pt}
\setlength{\belowcaptionskip}{4pt}
\includegraphics[angle=0, width=0.90 \columnwidth]{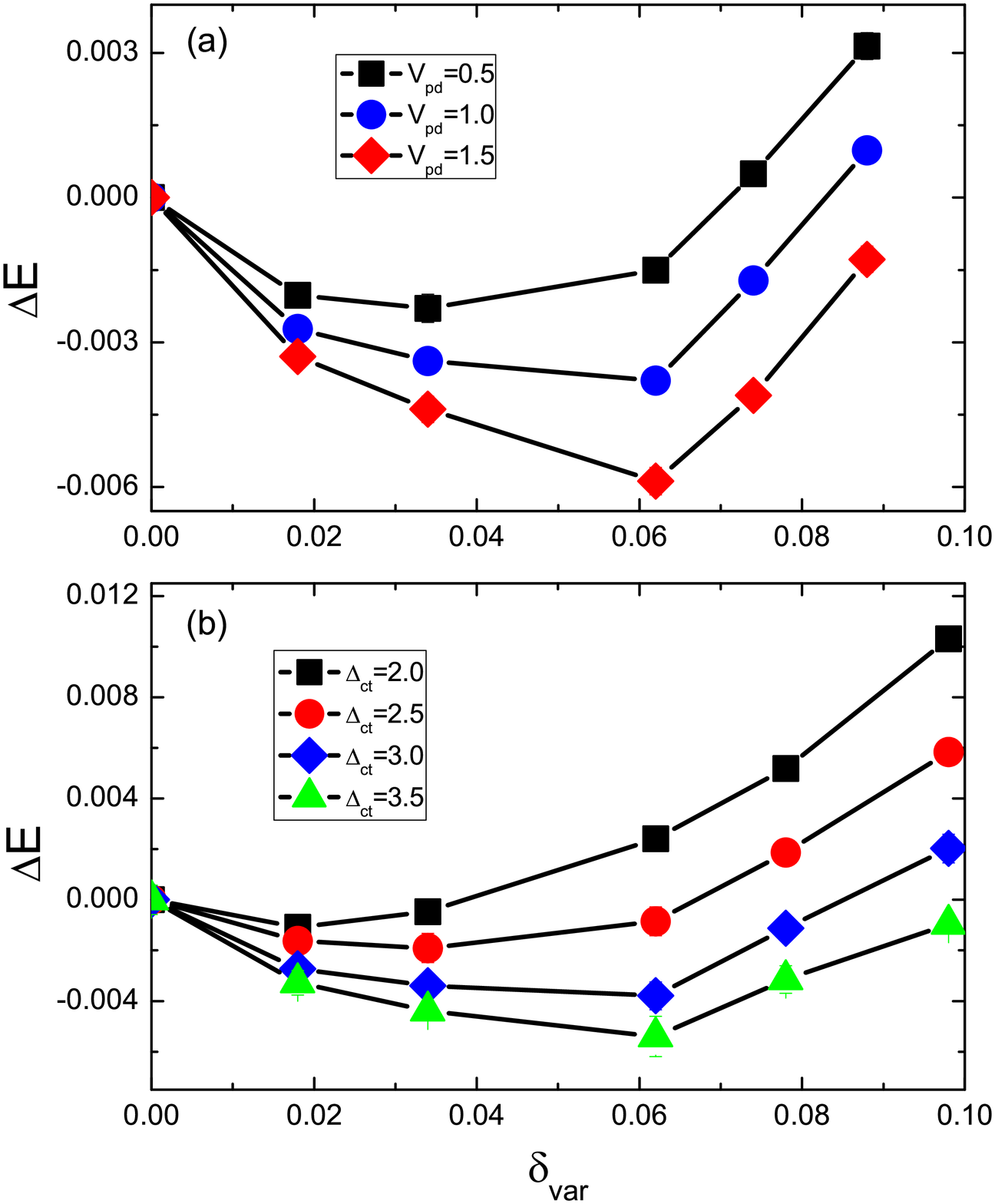}
\caption{(Color online) Dependence of total energy change on $\delta _{{\mathop{\rm
var}}}$ for different values of $V_{pd}$ (a) and ${\Delta _{ct}}$
(b) on the 20$\times$20 lattice and $n_h$=1.05.} \label{fig:2020dp105UdVpdDelta}
\end{figure}

\begin{figure}[htbp]
\centering \setlength{\abovecaptionskip}{2pt}
\setlength{\belowcaptionskip}{4pt}
\includegraphics[angle=0, width=0.98 \columnwidth]{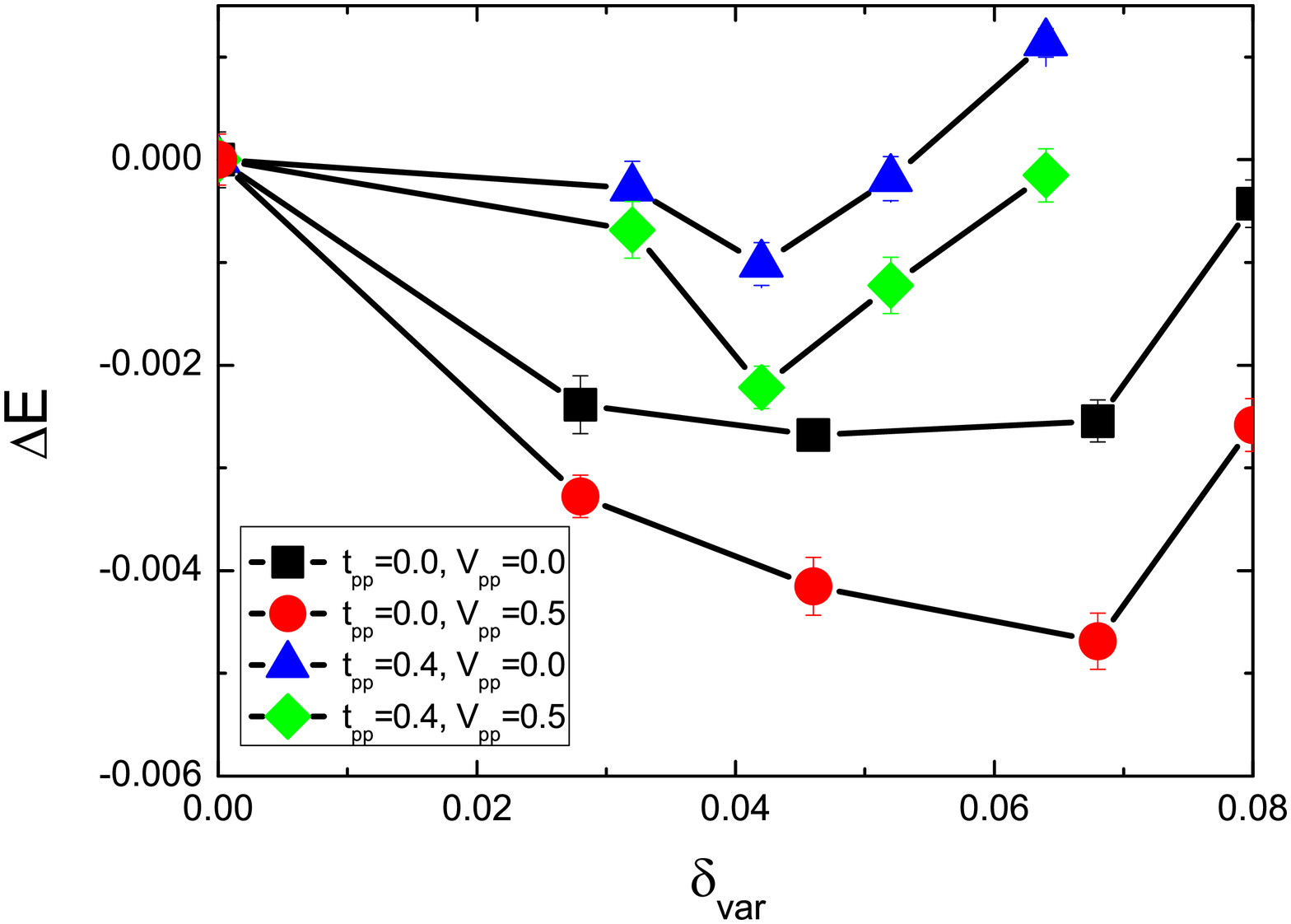}
\caption{(Color online) Total energy change as a function of $\delta _{{\mathop{\rm
var}}}$ for different values of $t_{pp}$ and $V_{pp}$. Lattice =
20$\times$20 and $n_h$=1.07. } \label{fig:2020dp107tppVpp}
\end{figure}

To explore the parameter dependence of the $d$FSD state, we present the
change of total energy $\Delta E =\left[ {E\left({{\delta_{{\mathop{\rm var}} }}}
\right) - E(0)} \right]/N$ as a function of $\delta _{{\mathop{\rm var}}}$ for
different sets of parameters at a fixed doping ($n_{h}=1.05$) in
Figs.~\ref{fig:2020dp105Ud}-\ref{fig:2020dp107tppVpp}.
From Fig.\ref{fig:2020dp105Ud}, one sees that the on-site Coulomb
interaction on the copper sites, $U_d$, has a remarkable positive effect on
$d$FSD, manifested by a decrease of $\Delta E$ with increasing $U_d$
and a shift of minimum to larger $\delta_{{\mathop{\rm var}}}$.
This can be expected since $U_d$ makes
a dominant contribution to the forward scattering, which is considered
to be the origin of $d$FSD. As seen in Fig.~\ref{fig:2020dp105UdVpdDelta},
a decrease of $\Delta E$ with increasing $V_{pd}$ or
${\Delta _{ct}}$ indicates that both NN copper-oxygen repulsion and charge-transfer
energy have a positive effect on the formation of $d$FSD.
It is naturally to expect that when $V_{pd}$ or ${\Delta _{ct}}$
increases, more holes transfer from oxygen sites to copper sites,
resulting in a stronger forward scattering of carriers by $U_d$.

Once $t_{pp}$ turns on, it may enhance the delocalization of holes
on the oxygen sites, and have a strong negative effect on $d$FSD.
To examine if $d$FSD exists in the actual cuprates materials,
we compare the change of total energy $\Delta E$
for $t_{pp}=0.0$ and $t_{pp}=0.4$
in Fig.~\ref{fig:2020dp107tppVpp}. It is readily seen that although
the condensation energy at $t_{pp}$=0.4 is much smaller than the
case of $t_{pp}$=0.0, the $d$FSD state still survives.

The NNN oxygen-oxygen Coulomb repulsion $V_{pp}$ is expected to have a
positive effect on $d$FSD, since it favors an inhomogeneity in the
hole densities of the NNN oxygen sites. In the previous work
by Fischer et al.~\cite{Fischer2011}, a mean field analysis based
on the three-band model showed that the minimum value of $V_{pp}$ that needed 
for a system to enter a nematic phase is about 1.2, which is much higher
than the realistic value. Here, in Fig.\ref{fig:2020dp107tppVpp}, we
show that the electronic nematicity has already shown up at
$V_{pp}$=0, and becomes stronger if we introduce $V_{pp}$=0.5. Our
results demonstrate that a spontaneous $d$FSD can occur in actual
cuprates materials.

\begin{figure}[htbp]
\centering \setlength{\abovecaptionskip}{2pt}
\setlength{\belowcaptionskip}{4pt}
\includegraphics[angle=0, width=0.98 \columnwidth]{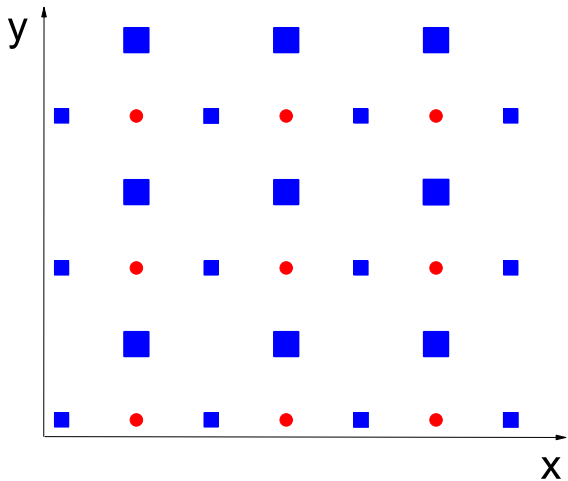}
\caption{(Color online) Distribution of hole densities on the copper and oxygen sites.
The results are obtained for the 20$\times$20 lattice at $n_h$=1.05.
The length of blue squares
represents the hole density at O$_x$ and O$_y$ sites, red circles
indicate the locations of Cu sites.} \label{fig:2020dp105noccupy}
\end{figure}

As mentioned before, an STM investigation on Bi$_2$Sr$_2$CaCu$_2$O$_{8+\delta}$~\cite{Lawler2010}
showed that there exists a significant density difference in
the two oxygen sites within each unit cell. In Fig.\ref{fig:2020dp105noccupy}, we present the
hole densities on the copper and oxygen sites for the 20$\times$20 lattice, at
$n_h$=1.05. The inhomogeneous distribution of hole densities in the
$p_x$ and $p_y$ oxygen orbitals indicates that $d$FSD in the three-band
model naturally results in an IUC nematicity, which is in good
agreement with the results of the STM investigation.
In the investigation of high-$T_c$ cuprates, nematic order is usually
considered to be the result of melted stripes~\cite{Hinkov2008,Daou2010},
namely, arising from proliferation of dislocations, the topological defects
in the striped state~\cite{Fradkin2010}. However, although a
striped state may have a close relation to the nematic order,
the $d$FSD instability can also lead directly to an electronic nematic
state, as shown in Fig.\ref{fig:2020dp105noccupy}.

\section{CONCLUSION}

In summary, our VMC study on the three-band Hubbard model show that
there exists a $d$FSD instability in the realistic parameter region of
cuprates, which is weakened with increasing the hole doping density and
vanishes at $n_{h}\sim 1.15$. The resulting imbalance of hole densities
on the NNN oxygen sites confirms that $d$FSD is a possible origin for
the IUC nematic phenomena, which was firstly proposed by Lawler
{\it et al.}~\cite{Lawler2010}. Our results show a qualitative agreement 
with the experimental results~\cite{Hinkov2008,Daou2010,Lawler2010} and
are consistent with the SBMF calculations for the $t$-$J$ model~\cite{Yamase2000}.
An analysis of the order dependence of $d$FSD on various parameters
within the three-band model showed that $U_d$, $V_{pd}$, and ${\Delta _{ct}}$
have a remarkable positive effect on $d$FSD. They contribute to the formation
of $d$FSD in such a way that the forward scattering is enhanced either
by increasing $U_d$ or through transferring holes from oxygen to copper
sites as $V_{pd}$ (${\Delta _{ct}}$) becomes larger.
We found that $d$FSD survives when a finite oxygen-oxygen hopping integrals $t_{pp}$
is introduced, and if a finite $V_{pp}$ interaction exists, the $d$FSD
state will be stabler.

Recently, the experimental evidence for electronic anisotropy in
iron-based superconductors has been
accumulated~\cite{Chuang2010,Chu2010,Fernandes2010,Fisher2011,Ying2011,Yi2011}.
The situation in iron-based superconductors is much more complicate since the
lattice, spin, and orbital degrees of freedom all manifest
themselves in the nematic state~\cite{Hu2011}. We consider that our
VMC study for the cuprates can be generalized to iron-based superconductors and
may be helpful for clarifying the nematicity in these systems.

\acknowledgements

This work was supported by the NSFC of China under Grant No. 11074257 and
11274310. Z.B.H. was supported by NSFC under Grant
Nos. 10974047 and 11174072, and by SRFDP under Grant No.20104208110001.
Numerical calculations were performed in Center for Computational
Science of CASHIPS.

\bibliography{MyCollection}% Produces the bibliography via BibTeX.

\begin{thebibliography}{}

\bibitem{Kivelson1998}
S. A. Kivelson, E. Fradkin, and V. J. Emery, Nature (London) 393,
550 (1998)

\bibitem{Hinkov2008}
V. Hinkov, D. Haug, B. Fauqu¡äe, P. Bourges, Y. Sidis, A. Ivanov, C.
Bernhard, C. T. Lin, and B. Keimer, Science 319, 597 (2008).

\bibitem{Daou2010}
R. Daou, J. Chang, D. LeBoeuf, O. Cyr-Choiniere, F. Laliberte, N.
Doiron-Leyraud, B. J. Ramshaw, R. Liang, D. A. Bonn, W. N. Hardy,
and L. Taillefer, Nature (London) 463, 519 (2010).

\bibitem{Lawler2010}
M. Lawler, K. Fujita, L. Jhinwhan, A. R. Schmidt, Y. Kohsaka, Chung
Koo Kim, H. Eisaki, S. Uchida, J. C. Davis, J. P. Sethna, and Eun-Ah
Kim, Nature (London) 466, 347 (2010).

\bibitem{Yamase2000}
H. Yamase and H. Kohno, J. Phys. Soc. Jpn. 69, 2151 (2000).

\bibitem{Miyanaga2006}
A. Miyanaga and H. Yamase, Phys. Rev. B 73, 174513 (2006).

\bibitem{Edegger2006}
B. Edegger, V. N. Muthukumar, and C. Gros, Phys. Rev. B 74, 165109
(2006).

\bibitem{Halboth2000}
C. J. Halboth and W. Metzner, Phys. Rev. Lett. 85, 5162 (2000).

\bibitem{Valenzuela2001}
B. Valenzuela and M. A. H. Vozmediano, Phys. Rev. B 63, 153103
(2001).

\bibitem{Hankevych2002}
V. Hankevych, I. Grote, and F. Wegner, Phys. Rev. B 66, 094516
(2002).

\bibitem{Kampf2003}
A.  P.  Kampf  and  A. A.  Katanin,  Phys.  Rev.  B  67,  125104
(2003).

\bibitem{Neumayr2003}
A. Neumayr and W. Metzner, Phys. Rev. B  67, 035112 (2003)

\bibitem{Kivelson2004}
S. A. Kivelson, E. Fradkin, and T. H. Geballe, Phys. Rev. B 69,
144505 (2004).

\bibitem{Emery1987}
V. J. Emery, Phys. Rev. Lett. 58, 2794 (1987).

\bibitem{Fischer2011}
M. H. Fischer and E.-A. Kim, Phys. Rev. B 84, 144502 (2011).

\bibitem{Yamase2005}
H. Yamase, V. Oganesyan, and W. Metzner, Phys. Rev. B 72, 035114
(2005).

\bibitem{Yamase2007}
H.  Yamase  and  W.  Metzner,  Phys.  Rev.  B  75,  155117
(2007).

\bibitem{Gros1988}
C. Gros, Phys. Rev. B  38, 931 (1988).

\bibitem{Yokoyama1988}
H. Yokoyama and H. Shiba, J. Phys. Soc. Jpn.  57, 2482 (1988)

\bibitem{Yokoyama1996}
H. Yokoyama and M. Ogata, ibid.  65, 3615 (1996).

\bibitem{Yamaji1998}
K. Yamaji, T. Yanagisawa, T. Nakanishi, and  S. Koike, J. Phys. Soc.
Jpn. 304, 225 (1998).

\bibitem{Becca2000}
F. Becca, M. Capone, and S. Sorella, Phys. Rev. B  62, 12700 (2000).

\bibitem{Paramekanti2004}
A. Paramekanti, M. Randeria, and N. Trivedi, Phys. Rev. Lett. 87,
217002 (2001); Phys. Rev. B  70, 054504 (2004).

\bibitem{Ceperley1977}
D. Ceperley, G. V. Chester, and K. H. Kalos, Phys. Rev. B 16, 3081
(1977).

\bibitem{Umrigar1988}
C. J. Umrigar, K. G. Wilson, and J. W. Wilkins, Phys. Rev.
Lett. 60, 1719 (1988).

\bibitem{Hybertsen1989}
M. S. Hybertsen, M.Schl\"{u}ter, and N. E.Christensen, Phys. Rev. B,
39, 9028 (1989).

\bibitem{Yanagisawa2001}
T. Yanagisawa, S. Koike, and K. Yamaji, Phys. Rev. B 64, 184509 (2001).

\bibitem{Fradkin2010}
E. Fradkin, S. A. Kivelson, M. J. Lawler, J. P. Eisenstein, and A. P. Mackenzie, Annu. Rev. Condens. Matter Phys. 1, 153 (2010).

\bibitem{Chuang2010}
T.-M. Chuang, M. P. Allan, J. Lee, Y. Xie, N. Ni, S. L. Bud¡¯ko, G.
S. Boebinger, P. C. Canfield, and J. C. Davis, Science 327, 181
(2010).

\bibitem{Chu2010}
J.-H. Chu, J. G. Analytis, K. De Greve, P. L. McMahon, Z. Islam, Y.
Yamamoto, and I. R. Fisher, Science  329, 824 (2010).

\bibitem{Fernandes2010}
R. M. Fernandes, L. H. VanBebber, S. Bhattacharya, P. Chandra, V.
Keppens, D. Mandrus, M. A. McGuire, B. C. Sales, A. S. Sefat, and J.
Schmalian, Phys. Rev. Lett. 105, 157003 (2010).

\bibitem{Fisher2011}
I. R. Fisher, L. Degiorgi, and Z. X. Shen, Rep. Prog. Phys. 74,
124506 (2011).

\bibitem{Ying2011}
J. J. Ying,  X. F. Wang, T. Wu, Z. J. Xiang,  R. H. Liu, Y. J. Yan,
A. F. Wang, M. Zhang, G. J. Ye, P. Cheng, et al., Phys. Rev. Lett.
107, 067001 (2011).

\bibitem{Yi2011}
M. Yi, D. H. Lu, J.-H. Chu, J. G. Analytis, A. P. Sorini, A. F.
Kemper, B. Moritz, S.-K. Mo, R. G. Moore, M. Hashimoto, W.-S. Lee,
Z. Hussain, T. P. Devereaux, I. R. Fisher, and Z.-X. Shen, PNAS 108,
6878 (2011).

\bibitem{Hu2011}
J Hu, C Xu, arXiv:1112.2713, (2011)


\end{thebibliography}

\end{document}